\documentclass[%
 reprint,
 amsmath,amssymb,
 aps,
]{revtex4-2}

\usepackage{graphicx}
\usepackage{dcolumn}
\usepackage{bm}

\usepackage{amssymb,amsfonts,amsmath,graphicx,physics}

\usepackage{color}
\usepackage{slashed}

\usepackage{hyperref}

\newcommand{\be}{\begin{equation}}
\newcommand{\ee}{\end{equation}}
\newcommand{\bea}{\begin{eqnarray}}
\newcommand{\eea}{\end{eqnarray}}

\newcommand{\dis}[1]{\begin{equation}\begin{split}#1\end{split}\end{equation}}

\newcommand{\bfrac}[2]{\left(\frac{#1}{#2} \right)  }
\newcommand{\eq}[1]{Eq.~(\ref{#1})}

\newcommand\gev{\,{\rm GeV}}

\newcommand\kev{\,{\rm keV}}
\newcommand\unitev{\,{\rm eV}}
\newcommand\cm{\,{\rm cm}}

\newcommand\km{\,{\rm km}}
\newcommand\pc{\,{\rm pc}}

\newcommand\s{{\rm s}}

\begin{document}

\preprint{APS/}

\title{Cored Dark Matter halos in the  Cosmic Neutrino Background}

\author{Wonsub Cho}
 \email{sub526@skku.edu}
\author{Ki-Young Choi}%
 \email{kiyoungchoi@skku.edu}
\affiliation{%
 Department of Physics, Sungkyunkwan University,  16419 Korea
 }%

\author{Hee Jung Kim}
 \email{heejungkim@ibs.re.kr}
\affiliation{
Center for Theoretical Physics of the Universe, \\Institute for Basic Science (IBS), Daejeon 34126, Republic of Korea
}


\begin{abstract}
We study the impact of the interaction between DM and the cosmic neutrino background on the evolution of galactic dark matter halos.
The energy transfer from the neutrinos to the dark matter can heat the center of the galaxy and make it cored. This effect is efficient for the small galaxies such as the satellite galaxies of the Milky Way and we can put conservative constraint on the  non-relativistic elastic scattering cross section as  $\sigma_{\chi\nu}\lesssim 10^{-31}\cm^2$ for 0.1 keV dark matter and 0.1 eV neutrino.
\end{abstract}

\maketitle


\section{\label{sec:intro}Introduction }

Two of the most persistent questions in particle physics are unveiling the nature of neutrinos and dark matter (DM).
While the role of neutrinos and their interactions within the Standard Model (SM) is phenomenologically well understood, the origin of their mass and mass hierarchies are unknown~\cite{Zyla:2020zbs}.
The evidences for the existence of DM are very convincing, but  all of them relies on the gravitational effect due to the  dominance of DM in mass density.
There have been continuous efforts to probe the DM interactions with ordinary matter through traditional experiments/observations~\cite{Arcadi:2017kky,Roszkowski:2017nbc}, and many ideas are suggested in new and creative ways.
In particular, the existence of interactions between neutrinos and DM is an interesting possibility that is often realized in the extension of SM to explain the nature of neutrinos and DM~\cite{Proceedings:2019qno,Capozzi:2018bps,Choi:2019zxy}. 

One way to probe DM-neutrino interactions is looking into the propagation of neutrinos.
Scattering among DM and neutrino can result in the additional attenuation of neutrino flux and off-diagonal interactions modify the neutrino flavor oscillations.
Deviation from the standard prediction could imply the existence of such interactions or constrain the interaction strength.
The propagation of energetic neutrinos from distant astrophysical sources is one ideal example.
Since DM is the dominant mass content and spreads all over the Universe, the astrophysical propagation length of the energetic neutrinos will maximize the imprints of DM-neutrino interactions in their propagation~\cite{Barranco:2010xt,Reynoso:2016hjr,Pandey:2018wvh,Choi:2019ixb,Choi:2020ydp}.
Moreover, local DM structures like the Galactic halo can provide anisotropic signatures in the neutrino sky~\cite{Arguelles:2017atb}.

The DM-neutrino interaction can also leave imprints in gravitational clustering of DM.
For example, stronger DM-neutrino scattering in the early Universe transfer kinetic energy to DM more efficiently, which results in warmer DM.
The resultant warmness of DM suppress their gravitational clustering, and it is indirectly probed through the modification of the cosmic microwave background (CMB) angular power spectra~\cite{Escudero:2015yka,DiValentino:2017oaw,Diacoumis:2018ezi} and the matter power spectrum probed by, e.g., the Lyman-$\alpha$ forest observations~\cite{Viel:2013fqw,Wilkinson:2014ksa,Hooper:2021rjc}.
While these bounds probe the DM-neutrino interaction in the early Universe, i.e., at rather high energies and flux of neutrinos, little has been studied in discussing the impact on late-time structure formation such as structural evolution of galaxies.

In this Letter, we study the effects of the interaction of DM with the cosmic neutrino background (C$\nu$B) on the evolution of DM halos.
The frequent interaction among them transfer kinetic energy between each other, which leads to core formation of DM halos.
In the central region of Milky Way (MW) satellites, DM particles are heated from the collisions with the cosmic neutrinos and thus moves outward;
since the kinetic energy of a DM particle is smaller towards the center, the impact of heating is larger toward the center.
Such a migration reduces the DM density from the center of a halo and forms a uniform density core.
For more massive structures like the MW or galaxy clusters, the DM kinetic energy is much larger and the heating is inefficient or oppositely DM can transfer energy to neutrinos.
In this study, we find an upper bound on the  DM-neutrino scattering cross section in the non-relativistic regime as $\sigma_{\chi \nu} \lesssim 10^{-31} \cm^2$ for the mass of DM $m_\chi = 0.1 \kev$ and the mass of neutrino $m_\nu = 0.1 \unitev$.

In Sec.~\ref{gravo}, we summarize the gravothermal fluid method we employed in this study, and in Sec.~\ref{DMCNB}, we solve those equations with the heat transfer between the neutrino background and DM.
We present the results in Sec.~\ref{core}, and conclude in Sec.~\ref{con}.

\section{Gravothermal fluid description of dark matter halos}
\label{gravo}%

We study the impact of DM heating from the DM-C$\nu$B scattering on the structural evolution of DM halos.
We numerically follow the evolution of isolated DM halos while taking into account the heating effect.

We assume that the DM heating from the DM-C$\nu$B scattering is unimportant during the course of halo formation.
This amounts to two approximations.
First, we approximate the DM halos to be formed around the time $t_*$ that is identical to conventional collisionless cold dark matter (CDM) halos, i.e., $t_*$ is determined by the virial mass $M_{200}$ of a halo based on a spherical collapse model~\cite{Press:1973iz,GalacticDynamics}, but $t_*$ does not depend on the DM-$\nu$ scattering cross section.
This assumption raises several caveats for our analyses, which will be discussed in the Concluding Remarks.
Second, we approximate the initial halo density profile by the Navarro-Frenk-White (NFW) profile~\cite{Navarro:1995iw,Navarro:1996gj} given by $\rho_{\rm NFW}(r)=\rho_s[(r/r_s)(1+r/r_s)^2]$.
The characteristic scale radius (density) $r_s$ ($\rho_s$) can be estimated using the mass-concentration relations determined from cosmological $N$-body simulations~\cite{Dutton:2014xda}, 
\dis{
    	r_s &\simeq 1.25~\text{kpc} \left( \frac{M_{200}}{10^{9}~\text{M}_\odot} \right)^{0.44},\\
    	\rho_s & \simeq 0.019 \text{M}_\odot / \text{pc}^3 \left( \frac{10^{9}~\text{M}_\odot}{M_{200}} \right)^{0.24}.
    	\label{eq:rho_s}
}

We employ the gravothermal fluid method~\cite{Balberg:2001qg,Balberg:2002ue,Ahn:2004xt,Koda:2011yb,Pollack:2014rja,Essig:2018pzq,Nishikawa:2019lsc} to follow the evolution of the DM halo density profile.
In this method, DM particles are effectively considered as an ideal fluid, which is described by their mass density $\rho$, and fluid pressure $p$. We define the 1-dimensional velocity dispersion $w = \sqrt{p/\rho^2}$.
Assuming the isotropic pressure and the skewlessness of the DM velocity distribution~\cite{Ahn:2004xt}, the gravothermal fluid equations are given by
\dis{
	\begin{split}
  &\text{continuity equation : }  \pdv{\rho}{t} + \frac{1}{r^2} \pdv{r} (r^2 \rho V_r) =0,\\
    &\text{Poisson's equation : }\frac{1}{r^2}\pdv{r}\left(r^2\pdv{\Phi}{r}  \right) = 4\pi G \rho,\\
&\text{Euler equation : } \pdv{V_r}{t} + V_r \pdv{V_r}{r} = -\pdv{\Phi}{r} -\frac{1}{\rho} \pdv{(\rho w^2)}{r},\\
&\text{entropy  : } \frac{1}{r^2} \pdv{r}(r^2 V_r) + \frac{3}{w}\left[ \pdv{w}{t} + V_r \pdv{w}{r}\right] = \frac{1}{w^2}\frac{\delta q}{\delta t},
\label{eq:Gravo}
	\end{split}
}
where $G$ is the Newton's constant.
$V_r=\expval{v_r}$ is the averaged velocity of DM in the radial direction with its relation to the velocity dispersion $w$ as $w^2 = \langle (v_r  -  \langle v_r\rangle )^2 \rangle$, and $\Phi$ is the gravitational potential.
In the LHS of the entropy equation, $\delta q / \delta t$ is the rate of the specific heat injection, i.e., the change in kinetic energy per DM mass and per time, which will be defined in the next section.

In the limit of quasi-static hydrostatic equilibrium, we can neglect the LHS of the Euler equation~[\eq{eq:Gravo}] and rewrite the equations in the Lagrangian formulation as
\begin{equation}
	\begin{split}
	    &\text{mass conservation : } \frac{\partial M}{\partial r}=4\pi r^{2}\rho\,,\\
		&\text{hydrostatic equilibrium : } \frac{\partial\left(\rho w^{2}\right)}{\partial r}+\frac{GM\rho}{r^{2}}=0\,,\\
		&\text{entropy : }\left(\frac{\partial}{\partial t}\right)_{M}\ln\frac{w^{3}}{\rho}=\frac{1}{w^2} \frac{\delta q}{\delta t},
	\end{split}
	\label{eq:Gravo_Fluid}
\end{equation}
where $M(r,t)$ is the fluid mass enclosed within the radius $r$ of a DM halo at time $t$, satisfying $r^2 \partial_r \Phi = GM$, and $(\partial_t)_M =  \partial_t + V_r \partial_r$. 
The assumption of quasi-static hydrostatic equilibrium is a good approximation when the gravitational timescale $t_G=1/\sqrt{4\pi G \rho}\simeq 0.1 \,{\rm Gyr} \,(0.01\,{\rm M}_\odot\cdot\pc^{-3} /\rho)^{-1/2}$ is shorter than the heating timescale defined as the inverse of the RHS of the entropy equation~[\eq{eq:Gravo_Fluid}], i.e., $t_{\rm heat}\equiv w^2/(\delta q/\delta t)$~\cite{Kamada:2019wjo};
the explicit value of the heating timescale will be presented in the following section.

\section{Dark Matter heating by C$\nu$B}
\label{DMCNB}%

The cosmic  neutrinos have decoupled from the thermal plasma at the cosmic temperature around MeV and travel around with the momentum redshift due to the expansion of the Universe, which is similar to the CMB. 
In the standard Big Bang Universe, the number density of the background neutrino per one flavor is
$n_{\nu,0} = n_{\bar{\nu},0} = 56\,(1+z)^3 \, \cm^{-3}$ with redshift $z$, assuming negligible lepton asymmetry of neutrino. Then  the neutrino number density for one flavor is
\dis{
n_{\nu} = n_{\nu,0}  + n_{\bar{\nu},0} = 112 \,(1+z)^3 \, \cm^{-3},
}
where the redshift is related to the cosmic time as 
\dis{
(1+z)^3 = \bfrac{t_{\rm age}}{t}^2,
}
in the matter-dominated Universe and  $t_{\rm age}=13.8 \, {\rm Gyr}$ is the present age of the Universe.

The mean velocity of these non-relativistic neutrino background  is~\cite{Ringwald:2004np}
\dis{
\expval{v_\nu}\simeq 1.6 \times 10^3 \, (1+z) \bfrac{0.1 \unitev}{m_\nu} \km/{\rm s},
}
and  the average kinetic energy is
\dis{
T_\nu &=\frac12 m_\nu \expval{v_\nu^2} \\
&\simeq 1.4\times10^{-9} \,\kev\,  (1+z)^2\bfrac{0.1\unitev}{m_\nu},
\label{Tnu}
}
where we used $\expval{v_\nu^2}\simeq \expval{v_\nu}^2$.
This can be compared with the average kinetic energy of  DM in the galaxies,
\dis{
T_\chi &= \frac12m_\chi \expval{v_\chi^2} \\
&\simeq 1.7\times 10^{-10} \kev \bfrac{m_\chi}{0.1\kev}\bfrac{w}{10\km/\s}^2,
\label{Tchi}
} 
where $v_\chi= |{\bf v}_\chi|$, and  in the last equation we used  $ \expval{v_\chi^2}  = 3 w^2$ for DM in the isotropic distribution and the quasi-static equilibrium. 

The scatterings of DM with the C$\nu$B exchange kinetic energy between them.
We can average this over all the initial neutrino directions and final scattering angles to obtain the average energy gain of DM (or equivalently, energy loss of the neutrino), which is given by \eq{DeltaE} in Appendix~\ref{KEtransfer}.
Integrating over the velocity distribution of DM, we find the average energy gain (or loss) by DM is
\dis{
\Delta E &= \frac{2m_\nu m_\chi}{(m_\nu+m_\chi)^2} \left(  T_\nu - T_\chi   \right),\\
&\simeq  2.8 \times 10^{-12}\kev \bfrac{0.1\kev}{m_\chi}  \left( 1 - \frac{ T_\chi}{T_\nu} \right).
\label{aveDeltaE}
}
Here, the cosmic neutrino background plays the role as the heat reservoir for DM.
When $T_\nu > T_\chi$, DM particles are heated from the scattering with the cosmic neutrinos, which happens in the center of dwarf galaxies.
Conversely, in more massive galaxies or when the mass of DM becomes larger,  $T_\nu < T_\chi$, and DM particles lose energy.

\begin{figure}[t]
\begin{center}
\begin{tabular}{c} 
	\includegraphics[width=0.45\textwidth]{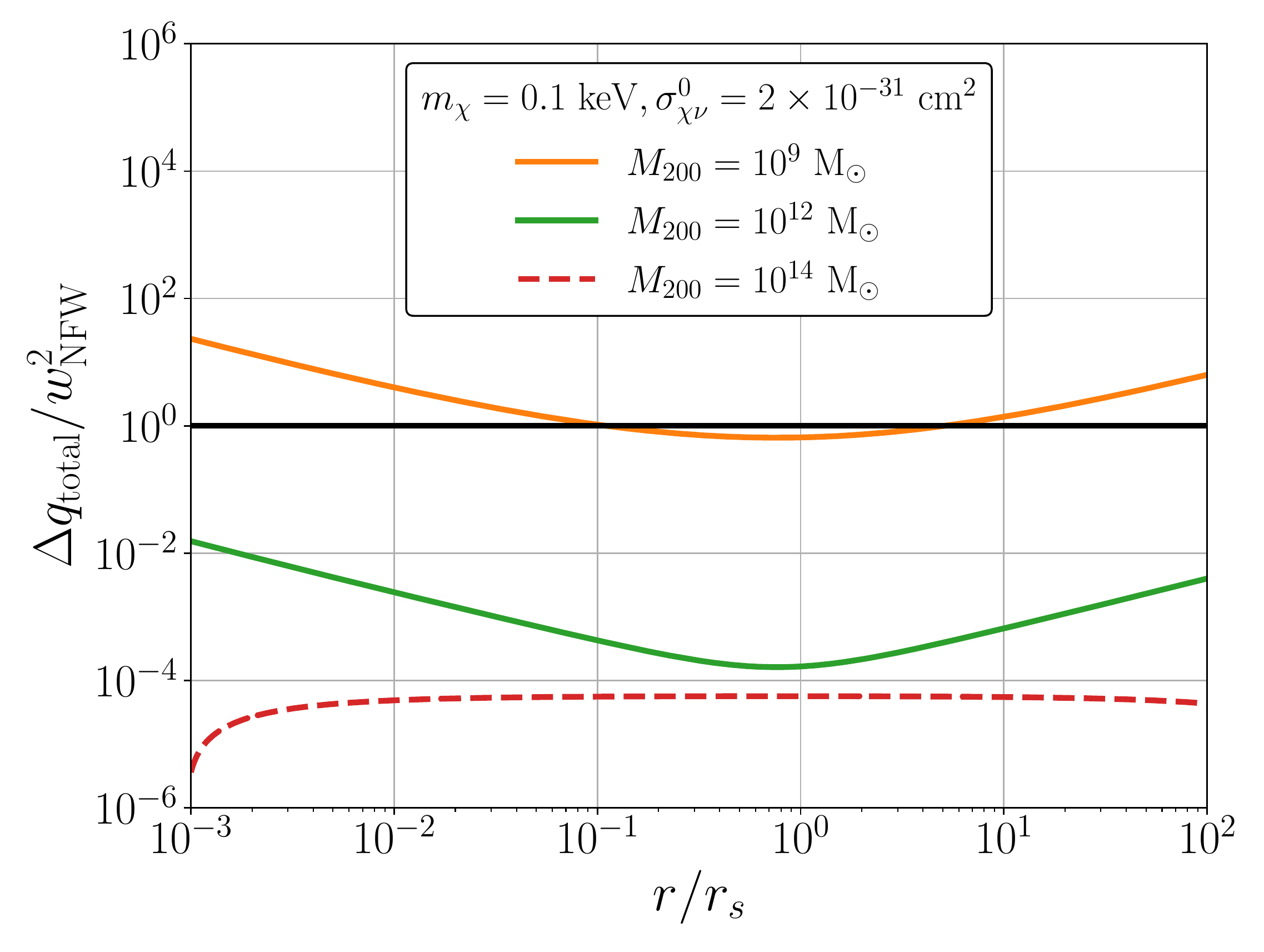}
	  \end{tabular}
\end{center}
        \caption{The ratio of the integrated heating and the initial NFW velocity dispersion $\Delta q_{\rm total}/w^2_{\rm NFW}$ at radius $r/r_s$ during the evolution by the time  $t_{\rm age}=13.8 \, {\rm Gyr}$ for $M_{200} = (10^9, 10^{12},10^{14} )~\text{M}_\odot$, each representing a MW satellite galaxy (orange), a MW-like host galaxy (green), and a galaxy cluster (red). Here we used $m_\chi = 0.1\kev$, $m_\nu=0.1\unitev$, and $\sigma_{\chi \nu} = 2\times 10^{-31}~\text{cm}^2$. The dashed line represents for negative $\Delta q_{\rm total}$.  }
        \label{fig:theat}
\end{figure}

The collision rate of a DM with neutrinos for the cross section $ \sigma_{\chi \nu} $ is given by
\dis{
t^{-1}_{\rm coll} &\equiv n_\nu \sigma_{\chi \nu} \expval{v_\nu} \\
&= \frac{5\times10^{-4}(1+z)^4 }{\rm \, Gyr}\bfrac{\sigma_{\chi\nu}}{10^{-30}\cm^2}\bfrac{0.1 \unitev}{m_\nu}.
}
It means that a DM particle is scattered by the background neutrinos more than once during the evolution of the galaxy if $\sigma_{\chi\nu}(1+z)^4 \gtrsim 10^{-27}\cm^2$ for $m_\nu =0.1\unitev$.
We note that the scattering cross section here is for the non-relativistic neutrinos and DM.
At higher energies, the scattering cross section should be suppressed to avoid other astrophysical and cosmological constraints~\cite{Choi:2019ixb}. For example, $\sigma_{\chi\nu}/m_\chi \lesssim 10^{-33} \cm^2/\gev$ for $E_\nu\sim 0.1\,{\rm keV}$ from the Lyman-$\alpha$ forest~\cite{Wilkinson:2014ksa}.

The heating timescale is estimated as 
\dis{
t_{\rm heat}^{-1} &= \frac{1}{w^2}\frac{\delta q}{\delta t}\\
 &= \frac{n_\nu \sigma_{\chi \nu} \expval{v_\nu}}{w^2} \frac{\Delta E}{m_\chi}\\
&  \simeq \frac{9\times 10^{-6}(1+z)^6}{\rm Gyr} \bfrac{10\km/{\rm s}}{w}^2 \\
&\quad\times \bfrac{0.1\kev}{m_\chi}^2\bfrac{\sigma_{\chi\nu}}{10^{-30}\cm^2}\bfrac{0.1\unitev}{m_\nu},
\label{qheat}
}
where we assumed $T_\chi  \ll T_\nu$ in the second equality.
Due to the redshift dependence of the number density and velocity of the C$\nu$B, the dominant contribution of the heat transfer occurs at earlier times.

The impact of the heating on halo structures can be conveniently understood by comparing the initial kinetic energy of DM particles with the integrated kinetic energy per DM mass.
The initial kinetic energy per DM mass is $\sim  w^2_{\rm NFW}$ where $w_{\rm NFW}$ is the 1-dimensional velocity dispersion that is in hydrostatic equilibrium with the initial NFW density profile~\cite{Lokas:2000mu}.
The integrated kinetic energy transfer per DM mass until the present is given by
\dis{
\Delta q_{\rm total}\equiv \int_{t_*}^{t_{\rm age}}\bfrac{\delta q}{\delta t} dt\,.
}
In Fig.~\ref{fig:theat}, we show the ratio $\Delta q_{\rm total}/w_{\rm NFW}^2$ for various halo masses $M_{200} = (10^9,10^{12},10^{14})\,\text{M}_\odot$, each representing a MW satellite galaxy (orange), a MW-like host galaxy (green), and a galaxy cluster (red).
The dashed curve represents negative $\Delta q_{\rm total}$.  
When $\Delta q_{\rm total}/w^2_{\rm NFW}>1$, the resultant velocity dispersion is enhanced from the initial one, i.e., $w^2_{\rm NFW}$, and the DM density at the given radius decreases from $\rho_{\rm NFW}(r)$, as will be discussed in the next section.
In massive galaxies like galaxy clusters, the energy transfer is insufficient  $\Delta q_{\rm total}/w^2_{\rm NFW}\ll1$ and there are no significant changes in the DM halo profiles.

\begin{figure*}
		\includegraphics[width=0.43\textwidth]{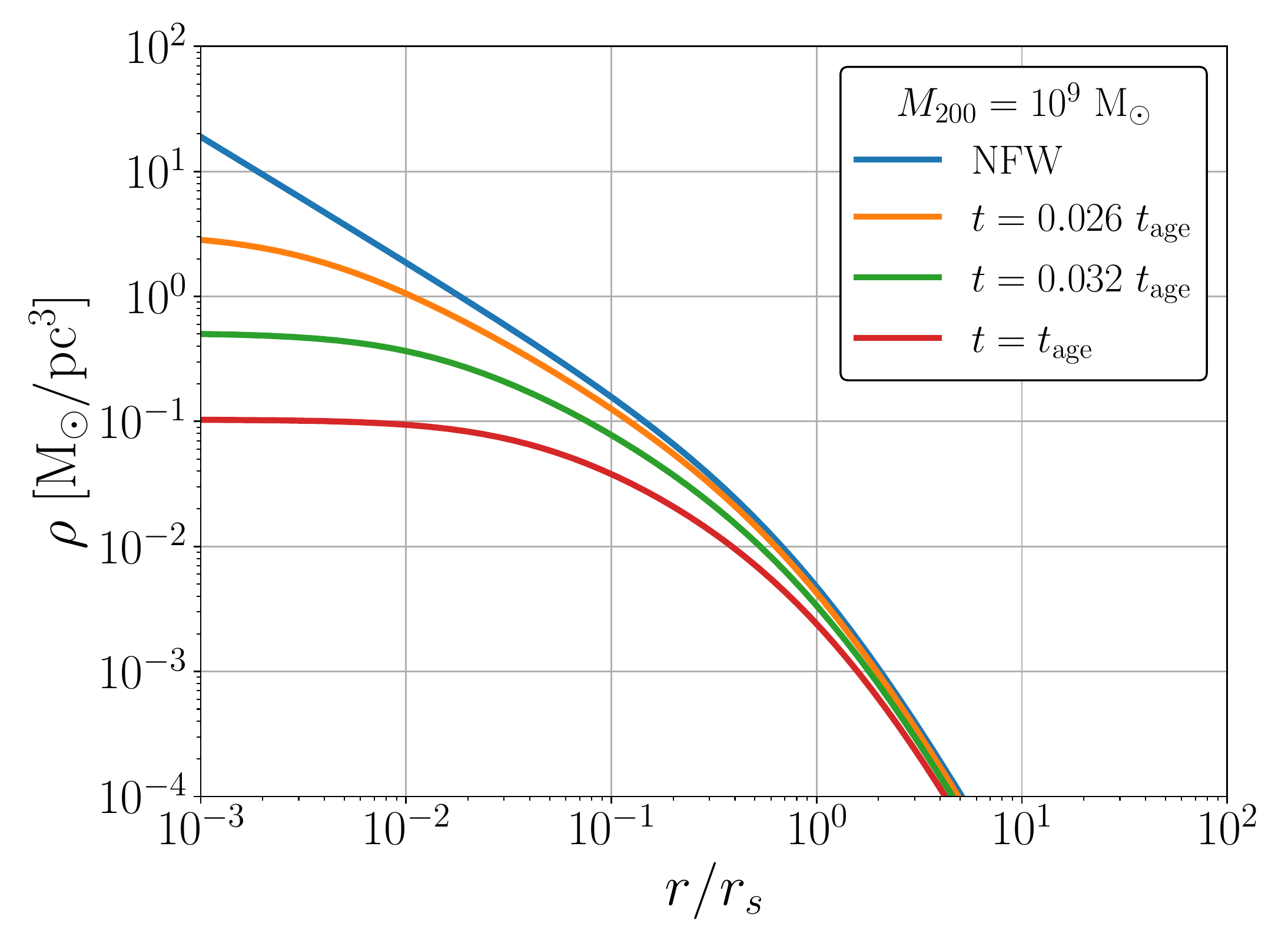}
		\includegraphics[width=0.42\textwidth]{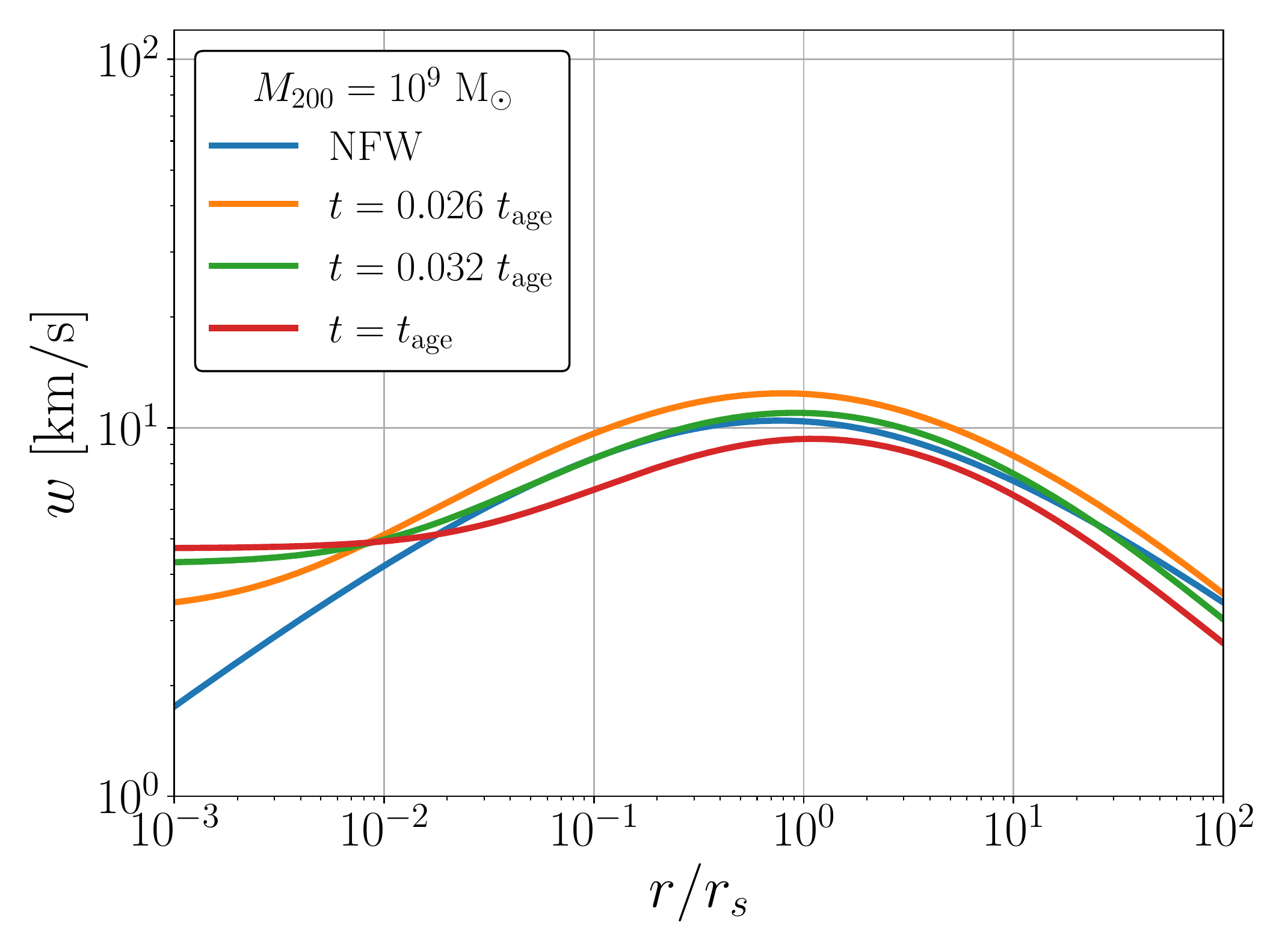}
        \caption{The evolution of DM mass density profile (left) and velocity dispersion (right) for  $M_{200} = 10^9~\text{M}_\odot$, such as a MW satellite galaxy.  Here we used $m_\chi = 0.1\kev$, $m_\nu=0.1\unitev$, $\sigma_{\chi \nu} = 2\times 10^{-31}~\text{cm}^2$ and $t_{\rm age}=13.8\,{\rm Gyr}$.}
        \label{fig:M9evolution}
\end{figure*}
%
\section{Core formation and DM-neutrino scattering}
\label{core}

We numerically follow the halo evolution by solving the gravothermal fluid equations in \eq{eq:Gravo_Fluid} with the heat transfer given by Eq.~\eqref{qheat} and Eqs.~\eqref{Tnu}-\eqref{aveDeltaE}.
In Fig.~\ref{fig:M9evolution}, we show the evolution of DM mass density profile (left) and velocity dispersion (right) for $M_{200} = 10^9~\text{M}_\odot$ at different epochs of  time, with the parameters  $m_\chi = 0.1\kev$, $m_\nu=0.1\unitev$, and $\sigma_{\chi \nu} = 2\times10^{-31}~\text{cm}^2$. 
With these benchmark parameters, the central density of the DM halo decreases and a core forms with the density $\rho_{\rm core}/\rho_s  \simeq 5$, while the velocity dispersion in the center grows to satisfy the hydrostatic equilibrium. 
The core density can be estimated from Fig.~\ref{fig:theat} with the core radius $r_{\rm core}$ where $(w^2_{\rm NFW}/\Delta q_{\rm total})|_{r_{\rm core}} =1$; $\rho_{\rm core}$ is estimated as
\dis{
\rho_{\rm core} \simeq 2\times\rho_{\rm NFW}(r=r_{\rm core}),
\label{eq:coreest}
}
where the numerical factor of 2 is determined empirically.
In the case of $\Delta q_{\rm total}/w^2_{\rm NFW} \gg 1$ for all radius,
the heating is insufficient and thus the central density is not affected.

\begin{figure*}
		\includegraphics[width=0.435\textwidth]{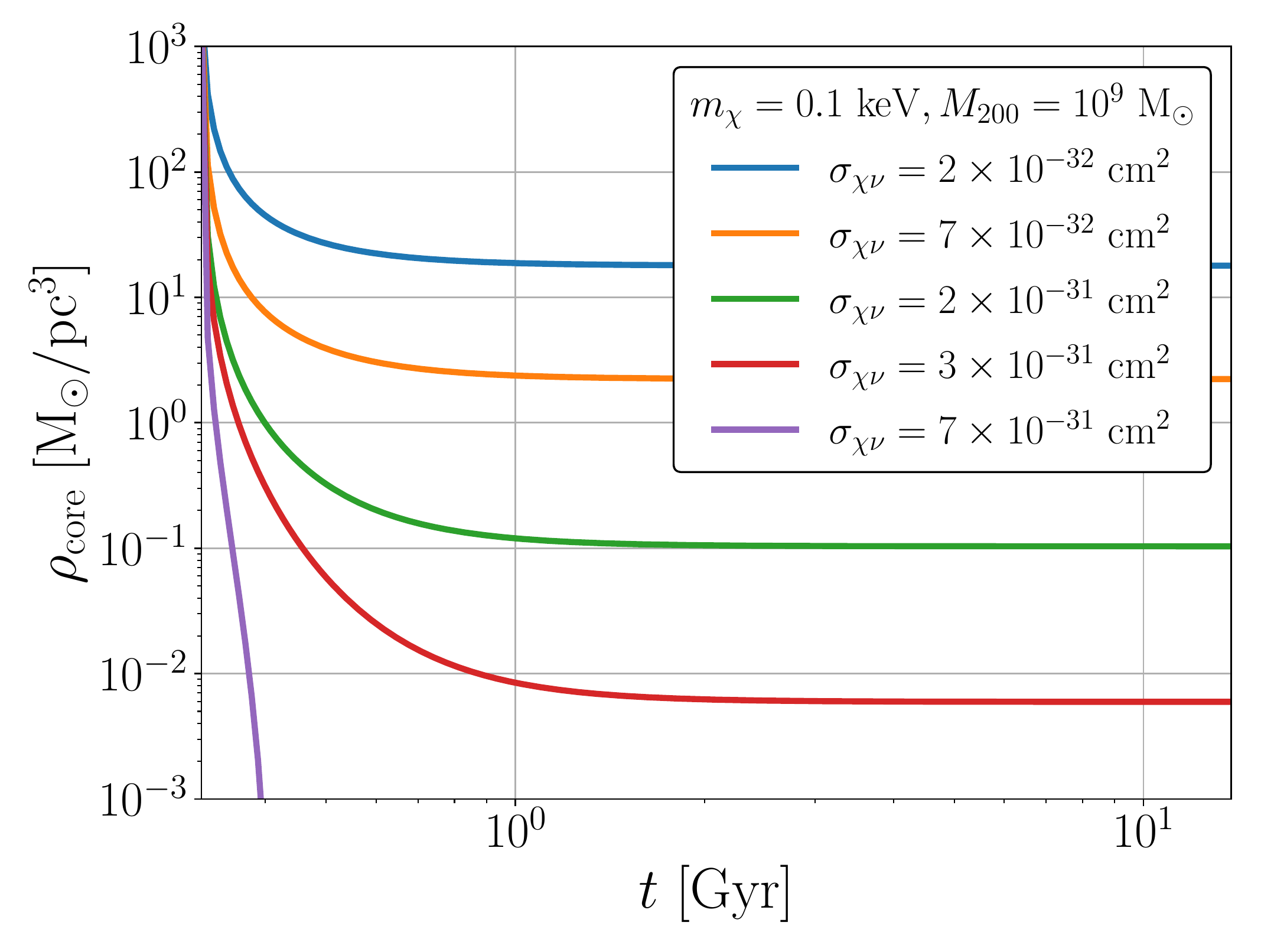}     
		\includegraphics[width=0.45\textwidth]{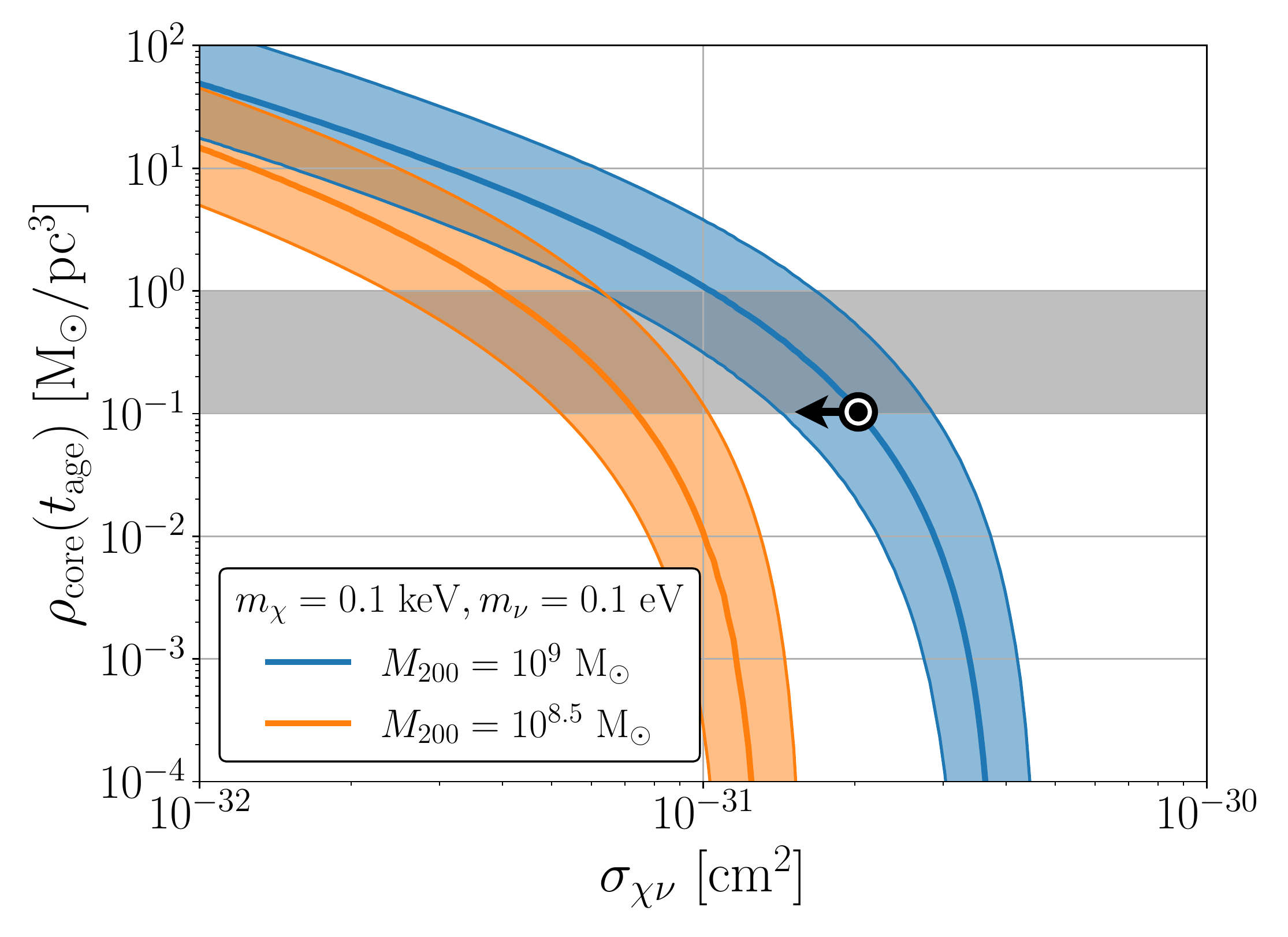}
		   \caption{Left: The time evolution of $\rho_{\rm core}$ for  different elastic scattering cross sections as shown in the figure
		    for $M_{200}=10^9 M_\odot$ with $m_\chi = 0.1\kev$, and $m_\nu=0.1\unitev$. Right: The dependence of $\rho_{\rm core}$ at time $t_{\rm age}$ on the cross section. The bands represent the $1\sigma$ spread in the concentration parameter, and the thick solid line represents the core density assuming a median value for the concentration parameter. We require the predicted core density to be larger than the observed one at present (gray shaded region, $\rho_{\rm c,obs}=0.1$-$1\,{\rm M_\odot/pc^3}$~\cite{Gilmore:2007fy}); for a MW satellite of mass $10^{9}\,{\rm M_\odot}$ with the median value for the concentration, we find that $\sigma_{\chi\nu}\lesssim 2\times 10^{-31}\,{\rm cm^2}$ for $m_{\chi}=0.1\,{\rm keV}$. }
        \label{fig:rho_core}
\end{figure*}

In the left panel of Fig.~\ref{fig:rho_core}, we show the time evolution of $\rho_{\rm core}$ for different elastic scattering cross sections  with $m_\chi = 0.1\kev$, and $m_\nu=0.1\unitev$ for a MW satellite galaxy ($10^9\,{\rm M}_{\odot}$). 
Due to the redshift dependence of the heating term, the core formation effect from the heating is efficient in the early time of the evolution, and later the core density is stabilized to a constant value. 
We can qualitatively understand the behavior of the time evolution from the entropy equation in \eq{eq:Gravo_Fluid};
there, if we ignore the change of $w$, the change of the DM density inside the uniform-core follows 
\dis{
-\left( \frac{1}{\rho}\pdv{\rho}{t}\right) _M \simeq \frac{1}{w^2} \frac{\delta q}{\delta t} \propto t^{-4}.
}
Therefore, $\rho_{\rm core}$ quickly decreases and approaches a constant value when the instantaneous heating timescale is longer compared to the age of the galaxy.

In the right panel of Fig.~\ref{fig:rho_core}, we show the values of $\rho_{\rm core}$ at the present for various values of the scattering cross section.
For $\sigma_{\chi\nu} \ll 10^{-31}\cm^2$, we see a certain power-law dependence of $\rho_{\rm core}$ with respect to $\sigma_{\chi\nu}$.
The scaling behavior can be understood by the following.
For such smaller values of $\sigma_{\chi\nu}$, the core radius $r_{\rm core}$ is smaller than the characteristic radius $r_s$, and thus we have an approximate relation $\rho_{\rm core}\propto 1/r_{\rm core}$
and $w^2_{\rm NFW}\propto r^{2/3}_{\rm core}$.
Meanwhile, remembering that $\Delta q _{\rm total}=w^2_{\rm NFW}$ at $r=r_{\rm core}$, one finds the scaling relation $\rho_{\rm core}\propto \sigma_{\chi\nu}^{-3/2}$.
For $\sigma_{\chi\nu} \gg 10^{-31}\cm^2$, $\Delta q_{\rm tot}>w_{\rm NFW}^2$ for all radius and the central density decreases exponentially.

For the cross sections larger than $\sigma_{\chi \nu}\gtrsim 2\times 10^{-31}\cm^2$, the core density of a MW satellite of mass $10^9\,{\rm M_\odot}$ (blue) becomes smaller than the observed central densities $\rho_{\rm c, obs}=0.1$-$1\,{\rm M_\odot/pc^3}$~\cite{Gilmore:2007fy}.
This argument gives the upper bound on the scattering cross section between DM and neutrino in the non-relativistic regime.
For example, if we require $\rho_{\rm core}\gtrsim 0.1\,{\rm M_\odot/pc^3}$ for a MW satellite of mass $10^9\,{\rm M_\odot}$, the upper bound is given as
\dis{
\sigma_{\chi\nu} \lesssim 2\times 10^{-31}\cm^2 \bfrac{m_\chi}{0.1\kev}^2 \bfrac{m_\nu}{0.1\unitev},
}
for non-relativistic DM and neutrino.
We remark that for DM heavier than $m_\chi \gg 1\kev$, the kinetic energy of DM can be larger than that of C$\nu$B and the DM halo is not affected.

The upper bound on $\sigma_{\chi \nu}$ is robust against the assumed concentration of a halo;
the bands in the right panel of Fig.~\ref{fig:rho_core} represent the $1\sigma$ spread in the concentration parameter, while the thick solid line represents the median value~\cite{Wechsler:2001cs,Dutton:2014xda}.
On the other hand, the upper bound is relatively more sensitive to the assumed halo mass.
If the assumed mass of a MW satellite is smaller, the heating would have greater impact since the typical kinetic energy of DM particles is smaller.
Thus, for a given cross section, smaller halos would exhibit smaller core density; compare the blue ($10^{9}\,{\rm M_\odot}$) and the orange ($10^{8.5}\,{\rm M_\odot}$) bands.
Nevertheless, we take $M_{\rm 200, infall}\simeq 10^{9}\,{\rm M_\odot}$ as a conservative benchmark mass of MW satellites where $M_{\rm 200, infall}$ is their masses prior to accretion onto the MW.
Note that our choice of halo mass gives a conservative upper bound on $\sigma_{\chi\nu}$ since progenitor halos of MW satellite galaxies lose their mass along the accretion.

%
\section{Concluding Remarks}
\label{con}
We have studied the impact of the interaction between DM and C$\nu$B on the DM halo evolution.
By employing the gravothermal fluid method, we have followed the evolution of the density profile and velocity dispersion of DM halos.
We find that the heat transfer from the C$\nu$B to DM flattens the initial cuspy density profile of DM and develops a uniform-density core.
This is because the heated DM from the DM-neutrino scattering migrates outwards to decrease DM density in the center.
The impact of the heating is stronger towards smaller-size halos, and we have used the observations on central densities of MW satellites to constrain the DM-neutrino scattering cross section in the non-relativistic regime as  $\sigma_{\chi\nu}\lesssim 10^{-31}\cm^2$ for 0.1 keV dark matter and 0.1 eV neutrino. This gives an independent bound on the DM and neutrino scattering at low energy compared to the previous constraints on the relativistic neutrinos.

In our study, we assumed that DM-C$\nu$B scattering do not affect the formation/virialization stage of halos.
Considering the importance of the heating towards early epochs, the initial virialized density profile of halos may not be the NFW profile and the intial central density could be more suppressed.
Furthermore, the possible warmness of DM from the DM-C$\nu$B scattering prior to the formation/virialization of a halo may delay the halo formation time.
In order to consistently take into account such effects, one needs to specify the velocity-dependence of the DM-neutrino cross section.
We leave the detailed study for future works.
Nevertheless, our study demonstrates that small-scale observations like the central densities of MW satellites can constrain the interaction between DM and C$\nu$B.

\begin{acknowledgments}
W.C. and K.-Y.C. were supported by the National Research Foundation of Korea (NRF) grant funded by the Korea government (MEST) (NRF-2019R1A2B5B01070181 and  NRF-2022R1A2C1005050). The work of H.K. is supported by IBS under the project code, IBS-R018-D1.
\end{acknowledgments}

\appendix

\section{Kinetic Energy Transfer}
\label{KEtransfer}
For non-relativistic and elastic collision of a  neutrino with velocity  ${\bf v}_1$ and  a DM with velocity ${\bf v}_2$, the kinetic energy transfer from the neutrino to DM is given by~\cite{Sigmund:2006}
\dis{
\Delta T =& \frac{1}{2} m_1 (v_1^2- {v_1'}^2 )  = m_1 {\bf V} \cdot({\bf u}_1 - { \bf u}_1'),
}
where ${\bf v}_1'$ and ${\bf v}_2'$ are velocities after collision and the  center-of-mass velocity is given by
\dis{
{\bf V}=  \frac{m_1 {\bf v}_1 + m_2 {\bf v}_2}{m_1+m_2}.
}
In the center-of-mass frame, the velocity of the neutrino before and after collision is
\dis{
{\bf u}_1 = {\bf v}_1  -  {\bf V}, \qquad {\bf u}_1' = {\bf v}_1'  -  {\bf V}.
}

Assuming that the neutrinos comes from all directions isotropically with the same magnitude with $v_1$, we can average the energy transfer to the DM which is moving with a fixed velocity ${\bf v_2}$ by integrating over the angles of the initial directions of the neutrinos and the final scattering angle. The averaged energy transfer from the neutrino to DM is given by
\dis{
\Delta E =& \frac{2m_1 m_2}{(m_1+m_2)^2} \left( T_1 - T_2 \right),
\label{DeltaE}
} 
where $T_1 = \frac12 m_1v_1^2$, and $T_2 = \frac12 m_2v_2^2$, respectively.


\bibliography{reference}

\begin{thebibliography}{36}%
\makeatletter
\providecommand \@ifxundefined [1]{%
 \@ifx{#1\undefined}
}%
\providecommand \@ifnum [1]{%
 \ifnum #1\expandafter \@firstoftwo
 \else \expandafter \@secondoftwo
 \fi
}%
\providecommand \@ifx [1]{%
 \ifx #1\expandafter \@firstoftwo
 \else \expandafter \@secondoftwo
 \fi
}%
\providecommand \natexlab [1]{#1}%
\providecommand \enquote  [1]{``#1''}%
\providecommand \bibnamefont  [1]{#1}%
\providecommand \bibfnamefont [1]{#1}%
\providecommand \citenamefont [1]{#1}%
\providecommand \href@noop [0]{\@secondoftwo}%
\providecommand \href [0]{\begingroup \@sanitize@url \@href}%
\providecommand \@href[1]{\@@startlink{#1}\@@href}%
\providecommand \@@href[1]{\endgroup#1\@@endlink}%
\providecommand \@sanitize@url [0]{\catcode `\\12\catcode `\$12\catcode
  `\&12\catcode `\#12\catcode `\^12\catcode `\_12\catcode `\%12\relax}%
\providecommand \@@startlink[1]{}%
\providecommand \@@endlink[0]{}%
\providecommand \url  [0]{\begingroup\@sanitize@url \@url }%
\providecommand \@url [1]{\endgroup\@href {#1}{\urlprefix }}%
\providecommand \urlprefix  [0]{URL }%
\providecommand \Eprint [0]{\href }%
\providecommand \doibase [0]{https://doi.org/}%
\providecommand \selectlanguage [0]{\@gobble}%
\providecommand \bibinfo  [0]{\@secondoftwo}%
\providecommand \bibfield  [0]{\@secondoftwo}%
\providecommand \translation [1]{[#1]}%
\providecommand \BibitemOpen [0]{}%
\providecommand \bibitemStop [0]{}%
\providecommand \bibitemNoStop [0]{.\EOS\space}%
\providecommand \EOS [0]{\spacefactor3000\relax}%
\providecommand \BibitemShut  [1]{\csname bibitem#1\endcsname}%
\let\auto@bib@innerbib\@empty
\bibitem [{\citenamefont {Zyla}\ \emph {et~al.}(2020)\citenamefont {Zyla} \emph
  {et~al.}}]{Zyla:2020zbs}%
  \BibitemOpen
  \bibfield  {author} {\bibinfo {author} {\bibfnamefont {P.}~\bibnamefont
  {Zyla}} \emph {et~al.} (\bibinfo {collaboration} {Particle Data Group}),\
  }\bibfield  {title} {\bibinfo {title} {{Review of Particle Physics}},\ }\href
  {https://doi.org/10.1093/ptep/ptaa104} {\bibfield  {journal} {\bibinfo
  {journal} {PTEP}\ }\textbf {\bibinfo {volume} {2020}},\ \bibinfo {pages}
  {083C01} (\bibinfo {year} {2020})}\BibitemShut {NoStop}%
\bibitem [{\citenamefont {Arcadi}\ \emph {et~al.}(2018)\citenamefont {Arcadi},
  \citenamefont {Dutra}, \citenamefont {Ghosh}, \citenamefont {Lindner},
  \citenamefont {Mambrini}, \citenamefont {Pierre}, \citenamefont {Profumo},\
  and\ \citenamefont {Queiroz}}]{Arcadi:2017kky}%
  \BibitemOpen
  \bibfield  {author} {\bibinfo {author} {\bibfnamefont {G.}~\bibnamefont
  {Arcadi}}, \bibinfo {author} {\bibfnamefont {M.}~\bibnamefont {Dutra}},
  \bibinfo {author} {\bibfnamefont {P.}~\bibnamefont {Ghosh}}, \bibinfo
  {author} {\bibfnamefont {M.}~\bibnamefont {Lindner}}, \bibinfo {author}
  {\bibfnamefont {Y.}~\bibnamefont {Mambrini}}, \bibinfo {author}
  {\bibfnamefont {M.}~\bibnamefont {Pierre}}, \bibinfo {author} {\bibfnamefont
  {S.}~\bibnamefont {Profumo}},\ and\ \bibinfo {author} {\bibfnamefont {F.~S.}\
  \bibnamefont {Queiroz}},\ }\bibfield  {title} {\bibinfo {title} {{The waning
  of the WIMP? A review of models, searches, and constraints}},\ }\href
  {https://doi.org/10.1140/epjc/s10052-018-5662-y} {\bibfield  {journal}
  {\bibinfo  {journal} {Eur. Phys. J. C}\ }\textbf {\bibinfo {volume} {78}},\
  \bibinfo {pages} {203} (\bibinfo {year} {2018})},\ \Eprint
  {https://arxiv.org/abs/1703.07364} {arXiv:1703.07364 [hep-ph]} \BibitemShut
  {NoStop}%
\bibitem [{\citenamefont {Roszkowski}\ \emph {et~al.}(2018)\citenamefont
  {Roszkowski}, \citenamefont {Sessolo},\ and\ \citenamefont
  {Trojanowski}}]{Roszkowski:2017nbc}%
  \BibitemOpen
  \bibfield  {author} {\bibinfo {author} {\bibfnamefont {L.}~\bibnamefont
  {Roszkowski}}, \bibinfo {author} {\bibfnamefont {E.~M.}\ \bibnamefont
  {Sessolo}},\ and\ \bibinfo {author} {\bibfnamefont {S.}~\bibnamefont
  {Trojanowski}},\ }\bibfield  {title} {\bibinfo {title} {{WIMP dark matter
  candidates and searches\textemdash{}current status and future prospects}},\
  }\href {https://doi.org/10.1088/1361-6633/aab913} {\bibfield  {journal}
  {\bibinfo  {journal} {Rept. Prog. Phys.}\ }\textbf {\bibinfo {volume} {81}},\
  \bibinfo {pages} {066201} (\bibinfo {year} {2018})},\ \Eprint
  {https://arxiv.org/abs/1707.06277} {arXiv:1707.06277 [hep-ph]} \BibitemShut
  {NoStop}%
\bibitem [{Pro(2019)}]{Proceedings:2019qno}%
  \BibitemOpen
  \href {https://doi.org/10.21468/SciPostPhysProc.2.001} {\emph {\bibinfo
  {title} {{Neutrino Non-Standard Interactions: A Status Report}}}},\
  Vol.~\bibinfo {volume} {2}\ (\bibinfo {year} {2019})\ \Eprint
  {https://arxiv.org/abs/1907.00991} {arXiv:1907.00991 [hep-ph]} \BibitemShut
  {NoStop}%
\bibitem [{\citenamefont {Capozzi}\ \emph {et~al.}(2018)\citenamefont
  {Capozzi}, \citenamefont {Shoemaker},\ and\ \citenamefont
  {Vecchi}}]{Capozzi:2018bps}%
  \BibitemOpen
  \bibfield  {author} {\bibinfo {author} {\bibfnamefont {F.}~\bibnamefont
  {Capozzi}}, \bibinfo {author} {\bibfnamefont {I.~M.}\ \bibnamefont
  {Shoemaker}},\ and\ \bibinfo {author} {\bibfnamefont {L.}~\bibnamefont
  {Vecchi}},\ }\bibfield  {title} {\bibinfo {title} {{Neutrino Oscillations in
  Dark Backgrounds}},\ }\href {https://doi.org/10.1088/1475-7516/2018/07/004}
  {\bibfield  {journal} {\bibinfo  {journal} {JCAP}\ }\textbf {\bibinfo
  {volume} {07}},\ \bibinfo {pages} {004}},\ \Eprint
  {https://arxiv.org/abs/1804.05117} {arXiv:1804.05117 [hep-ph]} \BibitemShut
  {NoStop}%
\bibitem [{\citenamefont {Choi}\ \emph
  {et~al.}(2020{\natexlab{a}})\citenamefont {Choi}, \citenamefont {Chun},\ and\
  \citenamefont {Kim}}]{Choi:2019zxy}%
  \BibitemOpen
  \bibfield  {author} {\bibinfo {author} {\bibfnamefont {K.-Y.}\ \bibnamefont
  {Choi}}, \bibinfo {author} {\bibfnamefont {E.~J.}\ \bibnamefont {Chun}},\
  and\ \bibinfo {author} {\bibfnamefont {J.}~\bibnamefont {Kim}},\ }\bibfield
  {title} {\bibinfo {title} {{Neutrino Oscillations in Dark Matter}},\ }\href
  {https://doi.org/10.1016/j.dark.2020.100606} {\bibfield  {journal} {\bibinfo
  {journal} {Phys. Dark Univ.}\ }\textbf {\bibinfo {volume} {30}},\ \bibinfo
  {pages} {100606} (\bibinfo {year} {2020}{\natexlab{a}})},\ \Eprint
  {https://arxiv.org/abs/1909.10478} {arXiv:1909.10478 [hep-ph]} \BibitemShut
  {NoStop}%
\bibitem [{\citenamefont {Barranco}\ \emph {et~al.}(2011)\citenamefont
  {Barranco}, \citenamefont {Miranda}, \citenamefont {Moura}, \citenamefont
  {Rashba},\ and\ \citenamefont {Rossi-Torres}}]{Barranco:2010xt}%
  \BibitemOpen
  \bibfield  {author} {\bibinfo {author} {\bibfnamefont {J.}~\bibnamefont
  {Barranco}}, \bibinfo {author} {\bibfnamefont {O.~G.}\ \bibnamefont
  {Miranda}}, \bibinfo {author} {\bibfnamefont {C.~A.}\ \bibnamefont {Moura}},
  \bibinfo {author} {\bibfnamefont {T.~I.}\ \bibnamefont {Rashba}},\ and\
  \bibinfo {author} {\bibfnamefont {F.}~\bibnamefont {Rossi-Torres}},\
  }\bibfield  {title} {\bibinfo {title} {{Confusing the extragalactic neutrino
  flux limit with a neutrino propagation limit}},\ }\href
  {https://doi.org/10.1088/1475-7516/2011/10/007} {\bibfield  {journal}
  {\bibinfo  {journal} {JCAP}\ }\textbf {\bibinfo {volume} {10}},\ \bibinfo
  {pages} {007}},\ \Eprint {https://arxiv.org/abs/1012.2476} {arXiv:1012.2476
  [astro-ph.CO]} \BibitemShut {NoStop}%
\bibitem [{\citenamefont {Reynoso}\ and\ \citenamefont
  {Sampayo}(2016)}]{Reynoso:2016hjr}%
  \BibitemOpen
  \bibfield  {author} {\bibinfo {author} {\bibfnamefont {M.~M.}\ \bibnamefont
  {Reynoso}}\ and\ \bibinfo {author} {\bibfnamefont {O.~A.}\ \bibnamefont
  {Sampayo}},\ }\bibfield  {title} {\bibinfo {title} {{Propagation of
  high-energy neutrinos in a background of ultralight scalar dark matter}},\
  }\href {https://doi.org/10.1016/j.astropartphys.2016.05.004} {\bibfield
  {journal} {\bibinfo  {journal} {Astropart. Phys.}\ }\textbf {\bibinfo
  {volume} {82}},\ \bibinfo {pages} {10} (\bibinfo {year} {2016})},\ \Eprint
  {https://arxiv.org/abs/1605.09671} {arXiv:1605.09671 [hep-ph]} \BibitemShut
  {NoStop}%
\bibitem [{\citenamefont {Pandey}\ \emph {et~al.}(2019)\citenamefont {Pandey},
  \citenamefont {Karmakar},\ and\ \citenamefont {Rakshit}}]{Pandey:2018wvh}%
  \BibitemOpen
  \bibfield  {author} {\bibinfo {author} {\bibfnamefont {S.}~\bibnamefont
  {Pandey}}, \bibinfo {author} {\bibfnamefont {S.}~\bibnamefont {Karmakar}},\
  and\ \bibinfo {author} {\bibfnamefont {S.}~\bibnamefont {Rakshit}},\
  }\bibfield  {title} {\bibinfo {title} {{Interactions of astrophysical
  neutrinos with dark matter: a model building perspective}},\ }\href
  {https://doi.org/10.1007/JHEP11(2021)215} {\bibfield  {journal} {\bibinfo
  {journal} {JHEP}\ }\textbf {\bibinfo {volume} {01}},\ \bibinfo {pages}
  {095}},\ \bibinfo {note} {[Erratum: JHEP 11, 215 (2021)]},\ \Eprint
  {https://arxiv.org/abs/1810.04203} {arXiv:1810.04203 [hep-ph]} \BibitemShut
  {NoStop}%
\bibitem [{\citenamefont {Choi}\ \emph {et~al.}(2019)\citenamefont {Choi},
  \citenamefont {Kim},\ and\ \citenamefont {Rott}}]{Choi:2019ixb}%
  \BibitemOpen
  \bibfield  {author} {\bibinfo {author} {\bibfnamefont {K.-Y.}\ \bibnamefont
  {Choi}}, \bibinfo {author} {\bibfnamefont {J.}~\bibnamefont {Kim}},\ and\
  \bibinfo {author} {\bibfnamefont {C.}~\bibnamefont {Rott}},\ }\bibfield
  {title} {\bibinfo {title} {{Constraining dark matter-neutrino interactions
  with IceCube-170922A}},\ }\href {https://doi.org/10.1103/PhysRevD.99.083018}
  {\bibfield  {journal} {\bibinfo  {journal} {Phys. Rev. D}\ }\textbf {\bibinfo
  {volume} {99}},\ \bibinfo {pages} {083018} (\bibinfo {year} {2019})},\
  \Eprint {https://arxiv.org/abs/1903.03302} {arXiv:1903.03302 [astro-ph.CO]}
  \BibitemShut {NoStop}%
\bibitem [{\citenamefont {Choi}\ \emph
  {et~al.}(2020{\natexlab{b}})\citenamefont {Choi}, \citenamefont {Chun},\ and\
  \citenamefont {Kim}}]{Choi:2020ydp}%
  \BibitemOpen
  \bibfield  {author} {\bibinfo {author} {\bibfnamefont {K.-Y.}\ \bibnamefont
  {Choi}}, \bibinfo {author} {\bibfnamefont {E.~J.}\ \bibnamefont {Chun}},\
  and\ \bibinfo {author} {\bibfnamefont {J.}~\bibnamefont {Kim}},\ }\bibfield
  {title} {\bibinfo {title} {{Dispersion of neutrinos in a medium}},\
  }\href@noop {} {\  (\bibinfo {year} {2020}{\natexlab{b}})},\ \Eprint
  {https://arxiv.org/abs/2012.09474} {arXiv:2012.09474 [hep-ph]} \BibitemShut
  {NoStop}%
\bibitem [{\citenamefont {Arg\"uelles}\ \emph {et~al.}(2017)\citenamefont
  {Arg\"uelles}, \citenamefont {Kheirandish},\ and\ \citenamefont
  {Vincent}}]{Arguelles:2017atb}%
  \BibitemOpen
  \bibfield  {author} {\bibinfo {author} {\bibfnamefont {C.~A.}\ \bibnamefont
  {Arg\"uelles}}, \bibinfo {author} {\bibfnamefont {A.}~\bibnamefont
  {Kheirandish}},\ and\ \bibinfo {author} {\bibfnamefont {A.~C.}\ \bibnamefont
  {Vincent}},\ }\bibfield  {title} {\bibinfo {title} {{Imaging Galactic Dark
  Matter with High-Energy Cosmic Neutrinos}},\ }\href
  {https://doi.org/10.1103/PhysRevLett.119.201801} {\bibfield  {journal}
  {\bibinfo  {journal} {Phys. Rev. Lett.}\ }\textbf {\bibinfo {volume} {119}},\
  \bibinfo {pages} {201801} (\bibinfo {year} {2017})},\ \Eprint
  {https://arxiv.org/abs/1703.00451} {arXiv:1703.00451 [hep-ph]} \BibitemShut
  {NoStop}%
\bibitem [{\citenamefont {Escudero}\ \emph {et~al.}(2015)\citenamefont
  {Escudero}, \citenamefont {Mena}, \citenamefont {Vincent}, \citenamefont
  {Wilkinson},\ and\ \citenamefont {B\oe{}hm}}]{Escudero:2015yka}%
  \BibitemOpen
  \bibfield  {author} {\bibinfo {author} {\bibfnamefont {M.}~\bibnamefont
  {Escudero}}, \bibinfo {author} {\bibfnamefont {O.}~\bibnamefont {Mena}},
  \bibinfo {author} {\bibfnamefont {A.~C.}\ \bibnamefont {Vincent}}, \bibinfo
  {author} {\bibfnamefont {R.~J.}\ \bibnamefont {Wilkinson}},\ and\ \bibinfo
  {author} {\bibfnamefont {C.}~\bibnamefont {B\oe{}hm}},\ }\bibfield  {title}
  {\bibinfo {title} {{Exploring dark matter microphysics with galaxy
  surveys}},\ }\href {https://doi.org/10.1088/1475-7516/2015/9/034} {\bibfield
  {journal} {\bibinfo  {journal} {JCAP}\ }\textbf {\bibinfo {volume} {09}},\
  \bibinfo {pages} {034}},\ \Eprint {https://arxiv.org/abs/1505.06735}
  {arXiv:1505.06735 [astro-ph.CO]} \BibitemShut {NoStop}%
\bibitem [{\citenamefont {Di~Valentino}\ \emph {et~al.}(2018)\citenamefont
  {Di~Valentino}, \citenamefont {B\o{}ehm}, \citenamefont {Hivon},\ and\
  \citenamefont {Bouchet}}]{DiValentino:2017oaw}%
  \BibitemOpen
  \bibfield  {author} {\bibinfo {author} {\bibfnamefont {E.}~\bibnamefont
  {Di~Valentino}}, \bibinfo {author} {\bibfnamefont {C.}~\bibnamefont
  {B\o{}ehm}}, \bibinfo {author} {\bibfnamefont {E.}~\bibnamefont {Hivon}},\
  and\ \bibinfo {author} {\bibfnamefont {F.~R.}\ \bibnamefont {Bouchet}},\
  }\bibfield  {title} {\bibinfo {title} {{Reducing the $H_0$ and $\sigma_8$
  tensions with Dark Matter-neutrino interactions}},\ }\href
  {https://doi.org/10.1103/PhysRevD.97.043513} {\bibfield  {journal} {\bibinfo
  {journal} {Phys. Rev. D}\ }\textbf {\bibinfo {volume} {97}},\ \bibinfo
  {pages} {043513} (\bibinfo {year} {2018})},\ \Eprint
  {https://arxiv.org/abs/1710.02559} {arXiv:1710.02559 [astro-ph.CO]}
  \BibitemShut {NoStop}%
\bibitem [{\citenamefont {Diacoumis}\ and\ \citenamefont
  {Wong}(2019)}]{Diacoumis:2018ezi}%
  \BibitemOpen
  \bibfield  {author} {\bibinfo {author} {\bibfnamefont {J.~A.~D.}\
  \bibnamefont {Diacoumis}}\ and\ \bibinfo {author} {\bibfnamefont {Y.~Y.~Y.}\
  \bibnamefont {Wong}},\ }\bibfield  {title} {\bibinfo {title} {{On the prior
  dependence of cosmological constraints on some dark matter interactions}},\
  }\href {https://doi.org/10.1088/1475-7516/2019/05/025} {\bibfield  {journal}
  {\bibinfo  {journal} {JCAP}\ }\textbf {\bibinfo {volume} {05}},\ \bibinfo
  {pages} {025}},\ \Eprint {https://arxiv.org/abs/1811.11408} {arXiv:1811.11408
  [astro-ph.CO]} \BibitemShut {NoStop}%
\bibitem [{\citenamefont {Viel}\ \emph {et~al.}(2013)\citenamefont {Viel},
  \citenamefont {Becker}, \citenamefont {Bolton},\ and\ \citenamefont
  {Haehnelt}}]{Viel:2013fqw}%
  \BibitemOpen
  \bibfield  {author} {\bibinfo {author} {\bibfnamefont {M.}~\bibnamefont
  {Viel}}, \bibinfo {author} {\bibfnamefont {G.~D.}\ \bibnamefont {Becker}},
  \bibinfo {author} {\bibfnamefont {J.~S.}\ \bibnamefont {Bolton}},\ and\
  \bibinfo {author} {\bibfnamefont {M.~G.}\ \bibnamefont {Haehnelt}},\
  }\bibfield  {title} {\bibinfo {title} {{Warm dark matter as a solution to the
  small scale crisis: New constraints from high redshift
  Lyman-\ensuremath{\alpha} forest data}},\ }\href
  {https://doi.org/10.1103/PhysRevD.88.043502} {\bibfield  {journal} {\bibinfo
  {journal} {Phys. Rev. D}\ }\textbf {\bibinfo {volume} {88}},\ \bibinfo
  {pages} {043502} (\bibinfo {year} {2013})},\ \Eprint
  {https://arxiv.org/abs/1306.2314} {arXiv:1306.2314 [astro-ph.CO]}
  \BibitemShut {NoStop}%
\bibitem [{\citenamefont {Wilkinson}\ \emph {et~al.}(2014)\citenamefont
  {Wilkinson}, \citenamefont {Boehm},\ and\ \citenamefont
  {Lesgourgues}}]{Wilkinson:2014ksa}%
  \BibitemOpen
  \bibfield  {author} {\bibinfo {author} {\bibfnamefont {R.~J.}\ \bibnamefont
  {Wilkinson}}, \bibinfo {author} {\bibfnamefont {C.}~\bibnamefont {Boehm}},\
  and\ \bibinfo {author} {\bibfnamefont {J.}~\bibnamefont {Lesgourgues}},\
  }\bibfield  {title} {\bibinfo {title} {{Constraining Dark Matter-Neutrino
  Interactions using the CMB and Large-Scale Structure}},\ }\href
  {https://doi.org/10.1088/1475-7516/2014/05/011} {\bibfield  {journal}
  {\bibinfo  {journal} {JCAP}\ }\textbf {\bibinfo {volume} {05}},\ \bibinfo
  {pages} {011}},\ \Eprint {https://arxiv.org/abs/1401.7597} {arXiv:1401.7597
  [astro-ph.CO]} \BibitemShut {NoStop}%
\bibitem [{\citenamefont {Hooper}\ and\ \citenamefont
  {Lucca}(2021)}]{Hooper:2021rjc}%
  \BibitemOpen
  \bibfield  {author} {\bibinfo {author} {\bibfnamefont {D.~C.}\ \bibnamefont
  {Hooper}}\ and\ \bibinfo {author} {\bibfnamefont {M.}~\bibnamefont {Lucca}},\
  }\bibfield  {title} {\bibinfo {title} {{Hints of dark matter-neutrino
  interactions in Lyman-$\alpha$ data}},\ }\href@noop {} {\  (\bibinfo {year}
  {2021})},\ \Eprint {https://arxiv.org/abs/2110.04024} {arXiv:2110.04024
  [astro-ph.CO]} \BibitemShut {NoStop}%
\bibitem [{\citenamefont {Press}\ and\ \citenamefont
  {Schechter}(1974)}]{Press:1973iz}%
  \BibitemOpen
  \bibfield  {author} {\bibinfo {author} {\bibfnamefont {W.~H.}\ \bibnamefont
  {Press}}\ and\ \bibinfo {author} {\bibfnamefont {P.}~\bibnamefont
  {Schechter}},\ }\bibfield  {title} {\bibinfo {title} {{Formation of galaxies
  and clusters of galaxies by selfsimilar gravitational condensation}},\ }\href
  {https://doi.org/10.1086/152650} {\bibfield  {journal} {\bibinfo  {journal}
  {Astrophys. J.}\ }\textbf {\bibinfo {volume} {187}},\ \bibinfo {pages} {425}
  (\bibinfo {year} {1974})}\BibitemShut {NoStop}%
\bibitem [{\citenamefont {Binney}\ and\ \citenamefont
  {Tremaine}(2011)}]{GalacticDynamics}%
  \BibitemOpen
  \bibfield  {author} {\bibinfo {author} {\bibfnamefont {J.}~\bibnamefont
  {Binney}}\ and\ \bibinfo {author} {\bibfnamefont {S.}~\bibnamefont
  {Tremaine}},\ }\href@noop {} {\emph {\bibinfo {title} {Galactic Dynamics:
  Second Edition}}},\ Princeton Series in Astrophysics\ (\bibinfo  {publisher}
  {Princeton University Press},\ \bibinfo {year} {2011})\BibitemShut {NoStop}%
\bibitem [{\citenamefont {Navarro}\ \emph {et~al.}(1996)\citenamefont
  {Navarro}, \citenamefont {Frenk},\ and\ \citenamefont
  {White}}]{Navarro:1995iw}%
  \BibitemOpen
  \bibfield  {author} {\bibinfo {author} {\bibfnamefont {J.~F.}\ \bibnamefont
  {Navarro}}, \bibinfo {author} {\bibfnamefont {C.~S.}\ \bibnamefont {Frenk}},\
  and\ \bibinfo {author} {\bibfnamefont {S.~D.~M.}\ \bibnamefont {White}},\
  }\bibfield  {title} {\bibinfo {title} {{The Structure of cold dark matter
  halos}},\ }\href {https://doi.org/10.1086/177173} {\bibfield  {journal}
  {\bibinfo  {journal} {Astrophys. J.}\ }\textbf {\bibinfo {volume} {462}},\
  \bibinfo {pages} {563} (\bibinfo {year} {1996})},\ \Eprint
  {https://arxiv.org/abs/astro-ph/9508025} {arXiv:astro-ph/9508025}
  \BibitemShut {NoStop}%
\bibitem [{\citenamefont {Navarro}\ \emph {et~al.}(1997)\citenamefont
  {Navarro}, \citenamefont {Frenk},\ and\ \citenamefont
  {White}}]{Navarro:1996gj}%
  \BibitemOpen
  \bibfield  {author} {\bibinfo {author} {\bibfnamefont {J.~F.}\ \bibnamefont
  {Navarro}}, \bibinfo {author} {\bibfnamefont {C.~S.}\ \bibnamefont {Frenk}},\
  and\ \bibinfo {author} {\bibfnamefont {S.~D.~M.}\ \bibnamefont {White}},\
  }\bibfield  {title} {\bibinfo {title} {{A Universal density profile from
  hierarchical clustering}},\ }\href {https://doi.org/10.1086/304888}
  {\bibfield  {journal} {\bibinfo  {journal} {Astrophys. J.}\ }\textbf
  {\bibinfo {volume} {490}},\ \bibinfo {pages} {493} (\bibinfo {year}
  {1997})},\ \Eprint {https://arxiv.org/abs/astro-ph/9611107}
  {arXiv:astro-ph/9611107} \BibitemShut {NoStop}%
\bibitem [{\citenamefont {Dutton}\ and\ \citenamefont
  {Macci\`o}(2014)}]{Dutton:2014xda}%
  \BibitemOpen
  \bibfield  {author} {\bibinfo {author} {\bibfnamefont {A.~A.}\ \bibnamefont
  {Dutton}}\ and\ \bibinfo {author} {\bibfnamefont {A.~V.}\ \bibnamefont
  {Macci\`o}},\ }\bibfield  {title} {\bibinfo {title} {{Cold dark matter haloes
  in the Planck era: evolution of structural parameters for Einasto and NFW
  profiles}},\ }\href {https://doi.org/10.1093/mnras/stu742} {\bibfield
  {journal} {\bibinfo  {journal} {Mon. Not. Roy. Astron. Soc.}\ }\textbf
  {\bibinfo {volume} {441}},\ \bibinfo {pages} {3359} (\bibinfo {year}
  {2014})},\ \Eprint {https://arxiv.org/abs/1402.7073} {arXiv:1402.7073
  [astro-ph.CO]} \BibitemShut {NoStop}%
\bibitem [{\citenamefont {Balberg}\ and\ \citenamefont
  {Shapiro}(2002)}]{Balberg:2001qg}%
  \BibitemOpen
  \bibfield  {author} {\bibinfo {author} {\bibfnamefont {S.}~\bibnamefont
  {Balberg}}\ and\ \bibinfo {author} {\bibfnamefont {S.~L.}\ \bibnamefont
  {Shapiro}},\ }\bibfield  {title} {\bibinfo {title} {{Gravothermal collapse of
  selfinteracting dark matter halos and the origin of massive black holes}},\
  }\href {https://doi.org/10.1103/PhysRevLett.88.101301} {\bibfield  {journal}
  {\bibinfo  {journal} {Phys. Rev. Lett.}\ }\textbf {\bibinfo {volume} {88}},\
  \bibinfo {pages} {101301} (\bibinfo {year} {2002})},\ \Eprint
  {https://arxiv.org/abs/astro-ph/0111176} {arXiv:astro-ph/0111176}
  \BibitemShut {NoStop}%
\bibitem [{\citenamefont {Balberg}\ \emph {et~al.}(2002)\citenamefont
  {Balberg}, \citenamefont {Shapiro},\ and\ \citenamefont
  {Inagaki}}]{Balberg:2002ue}%
  \BibitemOpen
  \bibfield  {author} {\bibinfo {author} {\bibfnamefont {S.}~\bibnamefont
  {Balberg}}, \bibinfo {author} {\bibfnamefont {S.~L.}\ \bibnamefont
  {Shapiro}},\ and\ \bibinfo {author} {\bibfnamefont {S.}~\bibnamefont
  {Inagaki}},\ }\bibfield  {title} {\bibinfo {title} {{Selfinteracting dark
  matter halos and the gravothermal catastrophe}},\ }\href
  {https://doi.org/10.1086/339038} {\bibfield  {journal} {\bibinfo  {journal}
  {Astrophys. J.}\ }\textbf {\bibinfo {volume} {568}},\ \bibinfo {pages} {475}
  (\bibinfo {year} {2002})},\ \Eprint {https://arxiv.org/abs/astro-ph/0110561}
  {arXiv:astro-ph/0110561} \BibitemShut {NoStop}%
\bibitem [{\citenamefont {Ahn}\ and\ \citenamefont
  {Shapiro}(2005)}]{Ahn:2004xt}%
  \BibitemOpen
  \bibfield  {author} {\bibinfo {author} {\bibfnamefont {K.-J.}\ \bibnamefont
  {Ahn}}\ and\ \bibinfo {author} {\bibfnamefont {P.~R.}\ \bibnamefont
  {Shapiro}},\ }\bibfield  {title} {\bibinfo {title} {{Formation and evolution
  of the self-interacting dark matter halos}},\ }\href
  {https://doi.org/10.1111/j.1365-2966.2005.09492.x} {\bibfield  {journal}
  {\bibinfo  {journal} {Mon. Not. Roy. Astron. Soc.}\ }\textbf {\bibinfo
  {volume} {363}},\ \bibinfo {pages} {1092} (\bibinfo {year} {2005})},\ \Eprint
  {https://arxiv.org/abs/astro-ph/0412169} {arXiv:astro-ph/0412169}
  \BibitemShut {NoStop}%
\bibitem [{\citenamefont {Koda}\ and\ \citenamefont
  {Shapiro}(2011)}]{Koda:2011yb}%
  \BibitemOpen
  \bibfield  {author} {\bibinfo {author} {\bibfnamefont {J.}~\bibnamefont
  {Koda}}\ and\ \bibinfo {author} {\bibfnamefont {P.~R.}\ \bibnamefont
  {Shapiro}},\ }\bibfield  {title} {\bibinfo {title} {Gravothermal collapse of
  isolated self-interacting dark matter haloes: N-body simulation versus the
  fluid model},\ }\href {https://doi.org/10.1111/j.1365-2966.2011.18684.x}
  {\bibfield  {journal} {\bibinfo  {journal} {Monthly Notices of the Royal
  Astronomical Society}\ }\textbf {\bibinfo {volume} {415}},\ \bibinfo {pages}
  {1125–1137} (\bibinfo {year} {2011})}\BibitemShut {NoStop}%
\bibitem [{\citenamefont {Pollack}\ \emph {et~al.}(2015)\citenamefont
  {Pollack}, \citenamefont {Spergel},\ and\ \citenamefont
  {Steinhardt}}]{Pollack:2014rja}%
  \BibitemOpen
  \bibfield  {author} {\bibinfo {author} {\bibfnamefont {J.}~\bibnamefont
  {Pollack}}, \bibinfo {author} {\bibfnamefont {D.~N.}\ \bibnamefont
  {Spergel}},\ and\ \bibinfo {author} {\bibfnamefont {P.~J.}\ \bibnamefont
  {Steinhardt}},\ }\bibfield  {title} {\bibinfo {title} {{Supermassive Black
  Holes from Ultra-Strongly Self-Interacting Dark Matter}},\ }\href
  {https://doi.org/10.1088/0004-637X/804/2/131} {\bibfield  {journal} {\bibinfo
   {journal} {Astrophys. J.}\ }\textbf {\bibinfo {volume} {804}},\ \bibinfo
  {pages} {131} (\bibinfo {year} {2015})},\ \Eprint
  {https://arxiv.org/abs/1501.00017} {arXiv:1501.00017 [astro-ph.CO]}
  \BibitemShut {NoStop}%
\bibitem [{\citenamefont {Essig}\ \emph {et~al.}(2019)\citenamefont {Essig},
  \citenamefont {Mcdermott}, \citenamefont {Yu},\ and\ \citenamefont
  {Zhong}}]{Essig:2018pzq}%
  \BibitemOpen
  \bibfield  {author} {\bibinfo {author} {\bibfnamefont {R.}~\bibnamefont
  {Essig}}, \bibinfo {author} {\bibfnamefont {S.~D.}\ \bibnamefont
  {Mcdermott}}, \bibinfo {author} {\bibfnamefont {H.-B.}\ \bibnamefont {Yu}},\
  and\ \bibinfo {author} {\bibfnamefont {Y.-M.}\ \bibnamefont {Zhong}},\
  }\bibfield  {title} {\bibinfo {title} {{Constraining Dissipative Dark Matter
  Self-Interactions}},\ }\href {https://doi.org/10.1103/PhysRevLett.123.121102}
  {\bibfield  {journal} {\bibinfo  {journal} {Phys. Rev. Lett.}\ }\textbf
  {\bibinfo {volume} {123}},\ \bibinfo {pages} {121102} (\bibinfo {year}
  {2019})},\ \Eprint {https://arxiv.org/abs/1809.01144} {arXiv:1809.01144
  [hep-ph]} \BibitemShut {NoStop}%
\bibitem [{\citenamefont {Nishikawa}\ \emph {et~al.}(2020)\citenamefont
  {Nishikawa}, \citenamefont {Boddy},\ and\ \citenamefont
  {Kaplinghat}}]{Nishikawa:2019lsc}%
  \BibitemOpen
  \bibfield  {author} {\bibinfo {author} {\bibfnamefont {H.}~\bibnamefont
  {Nishikawa}}, \bibinfo {author} {\bibfnamefont {K.~K.}\ \bibnamefont
  {Boddy}},\ and\ \bibinfo {author} {\bibfnamefont {M.}~\bibnamefont
  {Kaplinghat}},\ }\bibfield  {title} {\bibinfo {title} {{Accelerated core
  collapse in tidally stripped self-interacting dark matter halos}},\ }\href
  {https://doi.org/10.1103/PhysRevD.101.063009} {\bibfield  {journal} {\bibinfo
   {journal} {Phys. Rev. D}\ }\textbf {\bibinfo {volume} {101}},\ \bibinfo
  {pages} {063009} (\bibinfo {year} {2020})},\ \Eprint
  {https://arxiv.org/abs/1901.00499} {arXiv:1901.00499 [astro-ph.GA]}
  \BibitemShut {NoStop}%
\bibitem [{\citenamefont {Kamada}\ and\ \citenamefont
  {Kim}(2020)}]{Kamada:2019wjo}%
  \BibitemOpen
  \bibfield  {author} {\bibinfo {author} {\bibfnamefont {A.}~\bibnamefont
  {Kamada}}\ and\ \bibinfo {author} {\bibfnamefont {H.~J.}\ \bibnamefont
  {Kim}},\ }\bibfield  {title} {\bibinfo {title} {{Escalating core formation
  with dark matter self-heating}},\ }\href
  {https://doi.org/10.1103/PhysRevD.102.043009} {\bibfield  {journal} {\bibinfo
   {journal} {Phys. Rev. D}\ }\textbf {\bibinfo {volume} {102}},\ \bibinfo
  {pages} {043009} (\bibinfo {year} {2020})},\ \Eprint
  {https://arxiv.org/abs/1911.09717} {arXiv:1911.09717 [hep-ph]} \BibitemShut
  {NoStop}%
\bibitem [{\citenamefont {Ringwald}\ and\ \citenamefont
  {Wong}(2004)}]{Ringwald:2004np}%
  \BibitemOpen
  \bibfield  {author} {\bibinfo {author} {\bibfnamefont {A.}~\bibnamefont
  {Ringwald}}\ and\ \bibinfo {author} {\bibfnamefont {Y.~Y.~Y.}\ \bibnamefont
  {Wong}},\ }\bibfield  {title} {\bibinfo {title} {{Gravitational clustering of
  relic neutrinos and implications for their detection}},\ }\href
  {https://doi.org/10.1088/1475-7516/2004/12/005} {\bibfield  {journal}
  {\bibinfo  {journal} {JCAP}\ }\textbf {\bibinfo {volume} {12}},\ \bibinfo
  {pages} {005}},\ \Eprint {https://arxiv.org/abs/hep-ph/0408241}
  {arXiv:hep-ph/0408241} \BibitemShut {NoStop}%
\bibitem [{\citenamefont {Lokas}\ and\ \citenamefont
  {Mamon}(2001)}]{Lokas:2000mu}%
  \BibitemOpen
  \bibfield  {author} {\bibinfo {author} {\bibfnamefont {E.~L.}\ \bibnamefont
  {Lokas}}\ and\ \bibinfo {author} {\bibfnamefont {G.~A.}\ \bibnamefont
  {Mamon}},\ }\bibfield  {title} {\bibinfo {title} {{Properties of spherical
  galaxies and clusters with an nfw density profile}},\ }\href
  {https://doi.org/10.1046/j.1365-8711.2001.04007.x} {\bibfield  {journal}
  {\bibinfo  {journal} {Mon. Not. Roy. Astron. Soc.}\ }\textbf {\bibinfo
  {volume} {321}},\ \bibinfo {pages} {155} (\bibinfo {year} {2001})},\ \Eprint
  {https://arxiv.org/abs/astro-ph/0002395} {arXiv:astro-ph/0002395}
  \BibitemShut {NoStop}%
\bibitem [{\citenamefont {Gilmore}\ \emph {et~al.}(2007)\citenamefont
  {Gilmore}, \citenamefont {Wilkinson}, \citenamefont {Wyse}, \citenamefont
  {Kleyna}, \citenamefont {Koch}, \citenamefont {Evans},\ and\ \citenamefont
  {Grebel}}]{Gilmore:2007fy}%
  \BibitemOpen
  \bibfield  {author} {\bibinfo {author} {\bibfnamefont {G.}~\bibnamefont
  {Gilmore}}, \bibinfo {author} {\bibfnamefont {M.~I.}\ \bibnamefont
  {Wilkinson}}, \bibinfo {author} {\bibfnamefont {R.~F.~G.}\ \bibnamefont
  {Wyse}}, \bibinfo {author} {\bibfnamefont {J.~T.}\ \bibnamefont {Kleyna}},
  \bibinfo {author} {\bibfnamefont {A.}~\bibnamefont {Koch}}, \bibinfo {author}
  {\bibfnamefont {N.~W.}\ \bibnamefont {Evans}},\ and\ \bibinfo {author}
  {\bibfnamefont {E.~K.}\ \bibnamefont {Grebel}},\ }\bibfield  {title}
  {\bibinfo {title} {{The Observed properties of Dark Matter on small spatial
  scales}},\ }\href {https://doi.org/10.1086/518025} {\bibfield  {journal}
  {\bibinfo  {journal} {Astrophys. J.}\ }\textbf {\bibinfo {volume} {663}},\
  \bibinfo {pages} {948} (\bibinfo {year} {2007})},\ \Eprint
  {https://arxiv.org/abs/astro-ph/0703308} {arXiv:astro-ph/0703308}
  \BibitemShut {NoStop}%
\bibitem [{\citenamefont {Wechsler}\ \emph {et~al.}(2002)\citenamefont
  {Wechsler}, \citenamefont {Bullock}, \citenamefont {Primack}, \citenamefont
  {Kravtsov},\ and\ \citenamefont {Dekel}}]{Wechsler:2001cs}%
  \BibitemOpen
  \bibfield  {author} {\bibinfo {author} {\bibfnamefont {R.~H.}\ \bibnamefont
  {Wechsler}}, \bibinfo {author} {\bibfnamefont {J.~S.}\ \bibnamefont
  {Bullock}}, \bibinfo {author} {\bibfnamefont {J.~R.}\ \bibnamefont
  {Primack}}, \bibinfo {author} {\bibfnamefont {A.~V.}\ \bibnamefont
  {Kravtsov}},\ and\ \bibinfo {author} {\bibfnamefont {A.}~\bibnamefont
  {Dekel}},\ }\bibfield  {title} {\bibinfo {title} {{Concentrations of dark
  halos from their assembly histories}},\ }\href
  {https://doi.org/10.1086/338765} {\bibfield  {journal} {\bibinfo  {journal}
  {Astrophys. J.}\ }\textbf {\bibinfo {volume} {568}},\ \bibinfo {pages} {52}
  (\bibinfo {year} {2002})},\ \Eprint {https://arxiv.org/abs/astro-ph/0108151}
  {arXiv:astro-ph/0108151} \BibitemShut {NoStop}%
\bibitem [{\citenamefont {Sigmund}(2006)}]{Sigmund:2006}%
  \BibitemOpen
  \bibfield  {author} {\bibinfo {author} {\bibfnamefont {P.}~\bibnamefont
  {Sigmund}},\ }\bibfield  {title} {\bibinfo {title} {{Particle Penetration and
  Radiation Effects}},\ }\bibfield  {journal} {\bibinfo  {journal} {Springer}\
  }\href {https://doi.org/https://doi.org/10.1007/3-540-31718-X}
  {https://doi.org/10.1007/3-540-31718-X} (\bibinfo {year} {2006})\BibitemShut
  {NoStop}%
\end{thebibliography}%

\end{document}